# Using Adjacent Data Retransmission to Improve the Transmission Efficiency of Multi-hop Relay Networks

Xuesong Liu, Jie Wu, *Member, IEEE*，Meng Zhou

*Abstract*–Data transmission systems are widely used in various aspects of life, industry, research and other fields. Transmission systems have different characteristics for different application scenarios. Many of them require multi-hop transmission to extend the transmission range, such as Wireless Mesh Network (WMN), Mobile Ad-hoc Network (MANET), Wireless Sensor Networks (WSN) and so on. The biggest problem of multi-hop transmission lies in that packet loss caused by bit error ratio (BER) in multiple transmission processes increases greatly, which makes end-to-end transmission reliability and bandwidth utilization decrease. This paper analyzes the impact of BER on bandwidth utilization in multi-hop relay transmission and proposes a simple and flexible transmission mechanism that is generally applicable to multi-hop transmission scenarios.

## I. INTRODUCTION

IN some data transmission systems, if the physical link between data source and target is too long or the communication quality is so poor that direct communication can't be performed, it is quite necessary to use relay stations to forward data. Relay transmission has the advantage of reducing the overall path loss between relay stations and data source station. A common solution for wireless multi-hop relay networks is to apply the Automatic Repeat-reQuest (ARQ) mechanisms to improve end-to-end transmission performance. In [1] a link layer concept for multi-hop reliable and efficient transmission is proposed. In [2] the author proposed an improved and efficient ARQ mechanism according to the 802.16 MMR system. In [3] a new high layer ACK mechanism is proposed for real-time video transmission in multi-hop wireless network, aiming to ensure the quality-of-service (QoS) of the. In some other multi-hop transmission scenarios, such as Internet of Things and large-scale data acquisition systems, they have different requirements on data reliability, network latency, physical links, and relay station performance. Therefore, an easy to implement and widely applicable multi-hop transmission scheme is necessary. For this reason, this paper proposes a simple and flexible multi-hop transmission scheme based on relay ACK.

## II. ANALYSIS OF MULTI-HOP RELAY NETWORKS

This article focuses on the general relay transmission, suitable for both wired and wireless transmission. As shown in Fig. 1, the data packet is sent from the source station SRC, passes through *n* relay stations P in turn, and finally reaches the DST station.

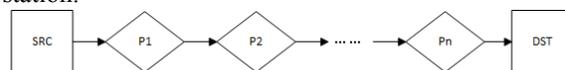

Fig. 1.  Multi-hop data transmission network model

When the physical link BER between adjacent stations is low, the data packet can be successfully transmitted from the SRC to the DST. When the link BER goes high, the throughput from the SRC to the DST is greatly reduced. Suppose that the link BER is ε, the number of transmission hops is $N$, and the data packet length is $L$ Byte. The packet loss rate between two adjacent stations is $W$, we get the following formula,

$$W = 1 - (1 - \varepsilon)^{8L} \qquad (1)$$

The packet loss rate from SRC to DST is denoted by P, then we get

$$P = 1 - (1 - W)^{N+1} \qquad (2)$$

According to (1) and (2), packet length $L = 1000$, get the relationship between BER and packet loss rate.

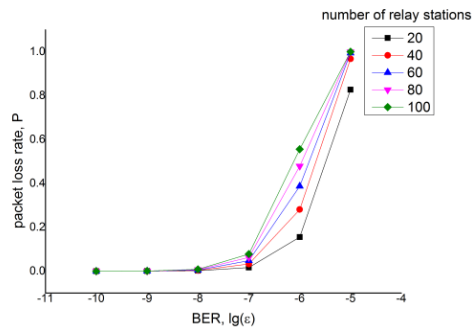

Fig. 2.  Relationship between BER and pack loss rate, L=1000

It can be seen from Fig. 2, when the BER is high, packet loss is very serious. When the number of relay stations increases, the packet loss rate also increases rapidly. When BER is $10^{-5}$, the packet loss rate of 100 relay stations exceeds 99.98%. In this

Manuscript received January 31, 2018. This project supported by NSFC (Grant No. 41574106), Major National Science and Technology Special Program of China (Grant No. 2017ZX05008-008-041), R&D of Key Instruments and Technologies for Deep Resources Prospecting (Grant No. ZDYZ2012-1-05-03).

The authors are with the Department of Modern Physics and State Key Laboratory of Particle Detection and Electronics, University of Science and Technology of China, Hefei, Anhui 230026, China (e-mail: wujie@ustc.edu.cn).

case, the data packet can no longer be transmitted from SRC to DST. Common multi-hop transmission protocol can't work.

## III. PROPOSED SCHEMES

In order to reduce the packet loss rate of multi-hop transmission and make full use of bandwidth, we propose a transmission mechanism suitable for multi-hop transmission system. Fig. 3 shows how the transmission mechanism works.

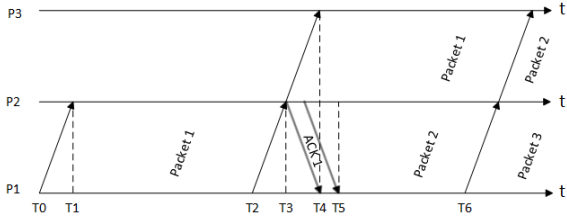

Fig. 3. The flow chart of our proposed scheme

As shown in Fig. 3, P1, P2 and P3 represent three adjacent relay stations. The data packet is transmitted from P1 to P3 direction. At time T0, P1 begins to transmit Packet1, which reaches P2 at T1. Then P2 begin to receive and check the content of Packet1. At time T2, P1 completes the sending of Packet1 and continues to send Packet2. At time T3, P2 completes Packet1 reception, and immediately sends an ACK to P1 if the check result is correct, and otherwise sends an N-ACK. At time T5, P1 receives the ACK from P2 and removes the buffered Packe1. If P1 has not received the ACK until T5, it resends Packet1 after Packet2 is sent. T5-T2 is the timeout period for P1 waiting for ACK.

$$T_{timeout} = 2RTT + 3T_{ACK} \quad (3)$$

In (3), RTT is round-trip time between two stations, and $T_{ACK}$ is the time of sending ACK packet. When $T_{timeout}$ is long, the relay station needs sufficient buffer space to temporarily store the packet. Due to the above transmission mechanism, the packet loss rate $W'$ of two adjacent stations decreases with the increase of transmission times.

$$W' = (8\varepsilon L)^n \quad (4)$$

The probability that an ACK is lost after multiple transmissions is far less than $W'$. In this case, if the packet loss due to buffer overflow is not considered, the packet loss rate from SRC to DST is

$$P' = (1 - (8\varepsilon L)^n)^N (1 - 8\varepsilon L) \quad (5)$$

## IV. SIMULATION

NS3 is a discrete-event network simulator for internet systems, we use it to simulate the case of multi-hop transmission. In the simulator, SRC transmission rate is set equal to link bandwidth, and there is no interval between packets. According to the receiving rate of DST, bandwidth utilization can be calculated. Fig. 4 simulates the relationship between transmissions times and bandwidth utilization. If relay station retransmits three times when packet loss occurs, and the number of relay stations is 100, the bandwidth utilization rate is 89.7% while the end-to-end transmission is less than 0.02%.

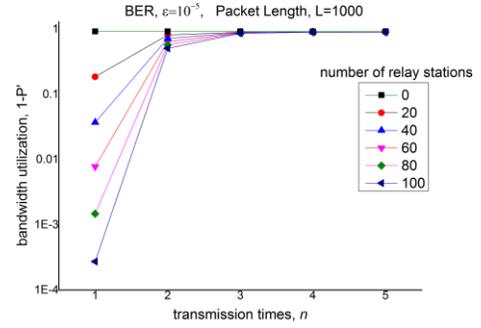

Fig. 4. Simulation results of the relationship between transmission times and bandwidth utilization

In some applications, relay station storage space is limited, some packets will inevitably be discarded due to buffer overflow. Further consider the impact of buffer size on bandwidth utilization, the simulation is shown in Fig. 5.

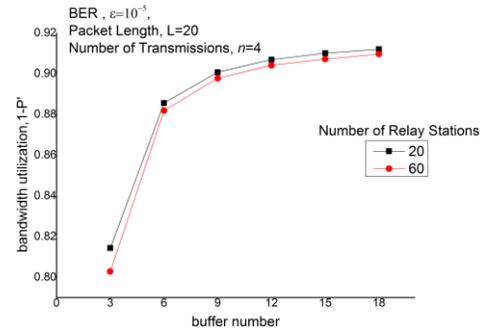

Fig. 5. Simulation results of the relationship between buffer number and bandwidth utilization

It can be seen from Fig. 6, if the buffer number is larger than 9, the bandwidth utilization ratio can exceed 90%, and then increasing the buffer size does not obviously increase the bandwidth utilization rate.

## V. CONCLUSION

The proposed transmission mechanism for multi-hop relay networks can make full use of link bandwidth at high BER. It also has the advantages of simplicity, flexibility, and small hardware requirements. In our future work, we plan to apply the mechanism in an FPGA-based wired data acquisition and transmission system that has a large number of relay stations.

REFERENCES

[1] H. Wiemann, M. Meyer, R. Ludwig and Chang Pae O, "A novel multi-hop ARQ concept," 2005 IEEE 61st Vehicular Technology Conference, 2005, pp. 3097-3101 Vol. 5.
[2] S. Y. Jeon, K. Y. Han, K. Suh and D. H. Cho, "An Efficient ARQ mechanism in Multi-Hop Relay Systems Based on IEEE 802.16 OFDMA," 2007 IEEE 66th Vehicular Technology Conference, Baltimore, MD, 2007, pp. 1649-1653.
[3] Z. Zhou, W. Jie, D. Xin, Z. ZhaoYu and T. Jun, "A new high layer ACK mechanism for real-time video transmission in multi-hop wireless network," 2016 International Conference on Internet of Things and Applications (IOTA), Pune, 2016, pp. 271-275.